\documentclass{article}

\PassOptionsToPackage{numbers, compress}{natbib}
\usepackage[final]{neurips_2025_ml4ps}

\usepackage[utf8]{inputenc} 
\usepackage[T1]{fontenc}    
\usepackage{hyperref}       
\usepackage{url}            
\usepackage{booktabs}       
\usepackage{amsfonts}       
\usepackage{nicefrac}       
\usepackage{microtype}      
\usepackage{xcolor}         
\usepackage{amsmath}
\usepackage{amssymb}
\usepackage{mathtools}
\usepackage{amsthm}
\usepackage{graphicx}

\usepackage{enumitem}
\usepackage{siunitx}
\usepackage{amsfonts}
\usepackage{nicefrac}
\usepackage{lipsum}
\usepackage{multirow}
\usepackage{color}
\usepackage{xspace}
\usepackage{soul}
\usepackage{listing}
\usepackage{amsmath}
\usepackage{mathtools}
\usepackage{multirow}
\usepackage{listings}
\usepackage{color}
\usepackage{here}
\usepackage{import}
\usepackage{bm}
\usepackage{adjustbox}
\usepackage{subcaption}
\usepackage{placeins}
\usepackage{siunitx}

\usepackage[capitalize,noabbrev]{cleveref}

\title{Sub-microsecond Transformers for Jet Tagging on FPGAs}

\author{
  \bfseries Lauri Laatu$^{1}$,
  Chang Sun$^{2}$,
  Arianna Cox$^{1}$,
  Abhijith Gandrakota$^{3}$, \\
  \bfseries Benedikt Maier$^{1}$, 
  Jennifer Ngadiuba$^{3}$, 
  Zhiqiang Que$^{1}$, \\
  \bfseries Wayne Luk$^{1}$,
  Maria Spiropulu$^{2}$,
  and Alexander Tapper$^{1}$
  \\[1.0em]
  \begin{tabular}{c}
    $^{1}$Imperial College London, United Kingdom \\
    $^{2}$California Institute of Technology, USA \\
    $^{3}$Fermilab, USA
  \end{tabular}
}

\begin{document}
\begin{flushright}
    FERMILAB-PUB-25-0779-CMS-LDRD
\end{flushright}

\maketitle

\begin{abstract}
    We present the first sub-microsecond transformer implementation on an FPGA achieving competitive performance for state-of-the-art high-energy physics benchmarks. Transformers have shown exceptional performance on multiple tasks in modern machine learning applications, including jet tagging at the CERN Large Hadron Collider (LHC). However, their computational complexity prohibits use in real-time applications, such as the hardware trigger system of the collider experiments up until now. In this work, we demonstrate the first application of transformers for jet tagging on FPGAs, achieving $\mathcal{O}(100)$ nanosecond latency with superior performance compared to alternative baseline models. We leverage high-granularity quantization and distributed arithmetic optimization to fit the entire transformer model on a single FPGA, achieving the required throughput and latency. Furthermore, we add multi-head attention and linear attention support to hls4ml, making our work accessible to the broader fast machine learning community. This work advances the next-generation trigger systems for the High Luminosity LHC, enabling the use of transformers for real-time applications in high-energy physics and beyond.
\end{abstract}

\section{Introduction}
\label{sec:introduction}

The CERN Large Hadron Collider (LHC)~\cite{lhc1995large} produces hundreds of terabytes of data per second from collisions at a frequency of 40\,MHz. To handle this vast data volume, two-level trigger systems are employed at experiments such as ATLAS~\cite{atlas} and CMS~\cite{CMS:2008xjf}. The first stage of this filter is the Level-1 trigger system (L1T), composed of hundreds of Field-Programmable Gate Arrays (FPGAs) that process data in real time and determine which collision events are of interest and thus are passed downstream for further analysis. As the data buffer size is limited, the data can only be retained for a short time, and the end-to-end latency of the L1T is limited to a few microseconds~\cite{cms-tdr-021,atlas-tdr-029}. Events not flagged by the L1T have to be permanently discarded, which makes the accuracy of the L1T critical.

The High-Luminosity LHC (HL-LHC)~\cite{hl-lhc} upgrade will increase the amount of simultaneous proton-proton collisions every \SI{25}{\nano\second} to as high as 140-200 from the current 40-80, which requires an extensive upgrade of the trigger systems. Extending the capability of the L1T to perform more informed decisions with deep learning based methods is therefore a key step toward the next-generation trigger systems. Research efforts~\cite{tgc,qkeras,qkeras-xtre-q,autoq,hls4ml,llp,compress-pipe,Bal:2023bvt} have focused on adapting machine learning models for FPGA deployment in real-time environments, showcasing their potential for ultra-low latency applications. Of particular relevance to improved L1T performance is the ability to identify (``tag'') jets, collimated sprays of particles that are key probes of the Standard Model and that appear in various new theories featuring hypothetical particles and forces. Ref.~\cite{qkeras} demonstrated that quantization-aware training (QAT) could enable jet tagging with quantized neural networks on FPGAs, achieving accuracy competitive to the full precision models while reducing resource usage by orders of magnitude. However, these models used high-level features that are typically not available at the L1T. In \cite{llgnn_old, llgnn,ds-fpga,mlpm-fpga, jedi-linear}, the feasibility of deploying practical jet-tagging neural networks on FPGAs with satisfactory performance and resource utilization for L1T is demonstrated with particle-level input. 

For offline data processing, state-of-the-art models for jet tagging include those based on Transformer architectures like the Particle Transformer~\cite{part} and the Lorentz-Equivariant Geometric Algebra Transformer~\cite{lagtr} as well as Graph Neural Networks (GNNs), such as ParticleNet~\cite{pnet} and PELICAN~\cite{pelican} which have shown exceptional performance in jet tagging tasks. However, such models are orders of magnitude too resource-intensive to be deployed on an FPGA in the real-time trigger system.


There have been attempts to deploy transformers on FPGAs for sub-microsecond latency applications. However, none of these approaches have yielded competitive performance compared to other methods, either only using the high-level features in~\cite{transformer_icl} or are unable to achieve the sub-microsecond latency in~\cite{transformer_uw}. In this work, we present the first successful implementation of transformers for jet tagging on FPGAs achieving $\mathcal{O}(100)$ nanosecond latency and better performance than current baseline models. The implementations we made for transformer support are also contributed back to the upstream \texttt{hls4ml} library~\cite{hls4ml_software}, making our work accessible to the broader fast machine learning community.

\section{Method}

\subsection{Model Architecture}

We adopt a simple encoder-only transformer architecture with vanilla multi-head attention (MHA)~\cite{transformer} with a single attention head. The input to the model is a sequence of particles, with a maximum number of (8, 16, 32, 64) particles sorted by $p_\mathrm{T}$, the transverse momentum. Each particle has three features: $p_\mathrm{T}$, $\eta$, and $\phi$. No positional encoding is added to the inputs, and the model used is a form of the Set Transformer~\cite{setattention}.

As the vanilla attention's computational complexity is $\mathcal{O}(n^2)$, where $n$ is the sequence length (which is the maximum number of particles in our case), we also implement the linear attention from Linformer~\cite{linformer} to improve model efficiency. In Linformer, the key and value vectors are projected to a lower dimension $k<n$. As such, the complexity is now $\mathcal{O}(k\cdot n)$. By selecting a $k$ decoupled from $n$, we are able to reduce the computational burden with approximated attention.



\subsection{Model Compression}

We adopt the High Granularity Quantization (HGQ)~\cite{hgq} for unified quantization and pruning of the networks. With gradient-based, per-parameter bitwidth optimization including zero width, the method performs loss-aware quantization and pruning simultaneously to reduce the model size. This approach significantly reduces the model size compared to layer-wise quantization methods such as those provided by QKeras~\cite{qkeras} and Brevitas~\cite{brevitas} while maintaining the accuracy of the original model.

HGQ works by introducing Effective Bit Operations (EBOPs) as an additional parameter where EBOPs is computed by summing the products of the bitwidths $ a_{bw} \cdot b_{bw} $ of all operations in the network. During training, the bitwidths are minimized as part of the loss. EBOPs has high correlation with the resource usage on FPGAs~\cite{hgq}, and it is possible to set a target EBOPs to achieve a desired size on hardware when training with HGQ. Value-wise heterogeneous quantization is also applied in the datapath, which breaks the otherwise permutation-invariant model to further reduce the firmware footprint.

We use \texttt{da4ml}~\cite{da4ml} to optimize the constant-matrix-vector multiplication (CMVM) operations used. The algorithm employed is a hybrid algorithm with both graph-based reduction and common subexpression elimination that transforms the CMVM operations into equivalent adder graphs. As the equivalence is exact, this optimization does not introduce any approximation error.

\section{Experiment}

We evaluate our methods using a common jet tagging dataset~\cite{hls4ml-dataset, dataset2}, a standard benchmark dataset used by the high-energy physics community~\cite{jedinet, llgnn_old,  llgnn, deepsetqkeras, mlpm-fpga, nas-fpga}. The dataset includes five classes of jets, each categorized by their originating particles: gluons (g), light quarks (q), W bosons (W), Z bosons (Z), and top quarks (t). This dataset contains 620,000 jets in the training set and 260,000 jets in the test set, each balanced between the classes. The FPGA target used for evaluation is the Xilinx XCU 250 chip. All experiments are performed using Vitis HLS and Vivado and the code used is part of \texttt{hls4ml} package.

\begin{figure}[h!]
    \begin{center}
        \includegraphics[width=0.67\linewidth]{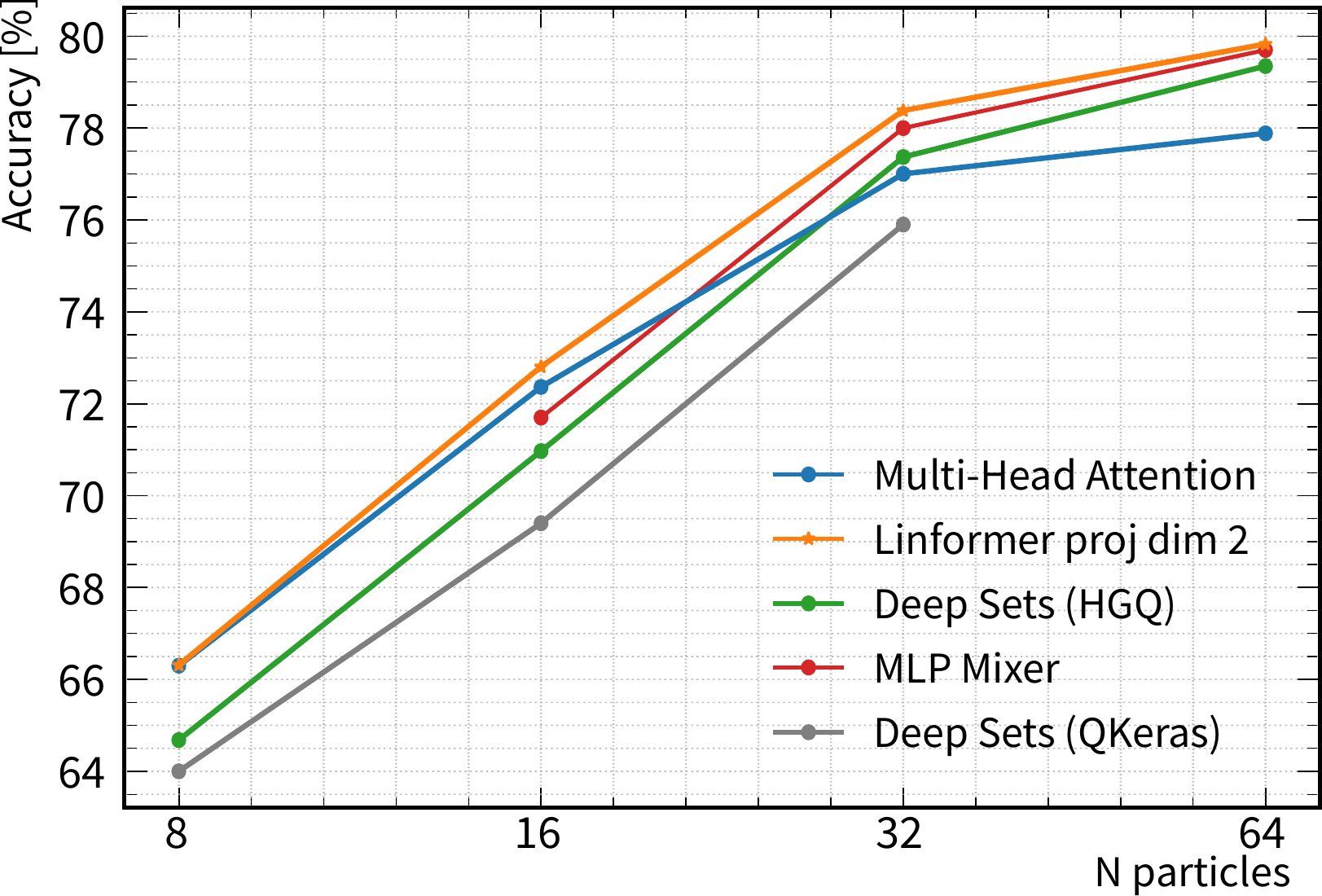}
    \end{center}
    \caption{Accuracy as a function of the maximum number of input particles. The decrease in accuracy for the full Transformer models compared to the other models at 64 particles is expected due to the fixed resource budget we enforced during training with EBOPs control.}
    \label{fig:acc}
\end{figure}

The transformer models are compared to methods from previous works: MLP Mixer-based jet tagging~\cite{mlpm-fpga}, Deep Sets with uniform quantization trained with QKeras~\cite{deepsetqkeras} and \cite{nas-fpga}, where the latter performed extensive neural architecture searches to find optimal tradeoffs between model accuracy and firmware performance. All models trained in this work have been trained with a target EBOPs of 350,000 - roughly one Super Logic Region of the target XCU 250 chip - with a proportion-integral-derivative (PID) controller over $\beta$, the regularization factor controlling the model size~\cite{hgq}, to enforce the target EBOPs.

\begin{figure}[h!]

    \centering

    \includegraphics[width=0.99\textwidth]{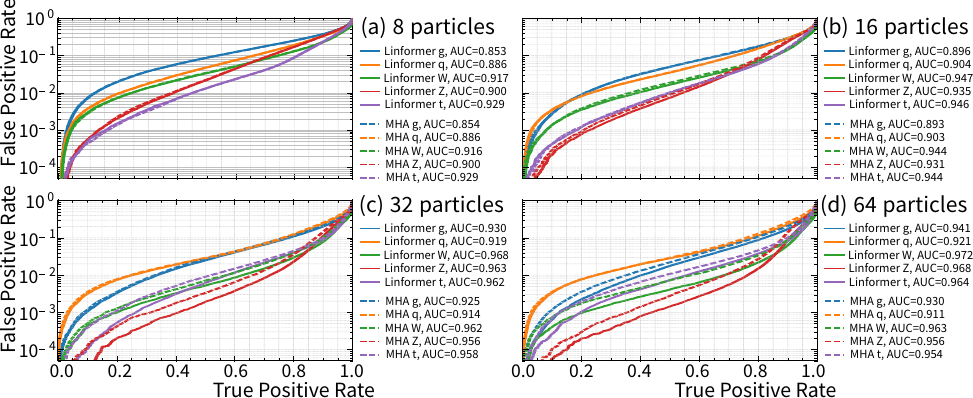}

    \caption{Receiver Operating Characteristic curves for different input lengths for MHA and Linformer.}
    \label{fig:roc2x2}

\end{figure}

The accuracy of the models as a function of number of input particles is shown in Figure~\ref{fig:acc}. The two attention methods both achieve state-of-the-art accuracy at 8 particles. However, as the input length increases, the quadratic scaling of the MHA model and the MHA layer requiring more FPGA resources, results in lower bitwidths across the network and the overall performance degrades compared to the other models. However, the Linformer models consistently outperform baseline methods across all input lengths. We also show the Receiver Operating Characteristic (ROC) curves in Figure~\ref{fig:roc2x2} for the efficiency per event type for different input lengths. All models exhibit a similar trend, with performance generally improving as the input length increases, although the rate of improvement varies between models and signals.

\begin{figure}[h!]

    \centering

    \includegraphics[width=0.99\textwidth]{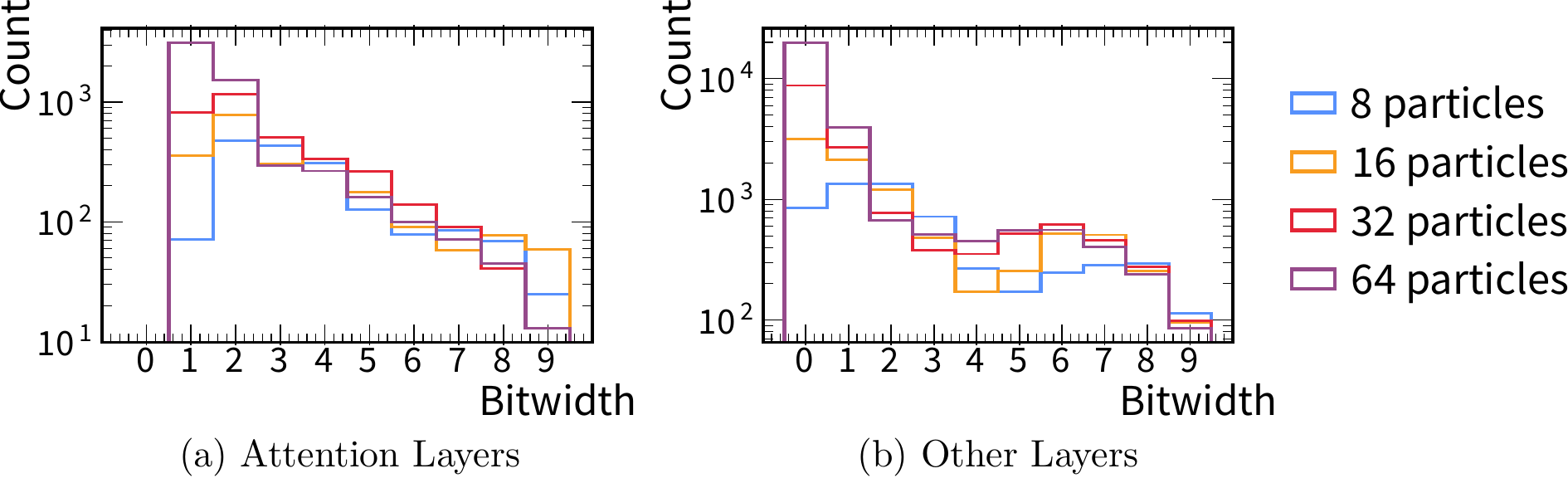}
    \caption{The distribution of the weight bitwidths of the obtained Linformer models. For the attention layers, the bitwidths are constrained to at least one bit, while other layers can adapt their bitwidth freely.}
    \label{fig:bitwidths}

\end{figure}

Bitwidths for the Linformer models are presented in Figure \ref{fig:bitwidths}. The bitwidths are divided into the attention layer which is constrained to at least one bit to disable pruning for training stability, and the other layers that do not have such constraints. As the target EBOPs is set to constant values, the model bitwidths have to adjust as the input length increases which explains the higher proportion of low bitwidths observed with larger input lengths.

The results of hardware synthesis are presented in Table \ref{tab:comparison}. The table shows that all transformer models are able to fulfill the $\mathcal{O}(100)$ nanosecond latency requirement. The attention block of the MHA model with 64 input particles is consistently collapsing over several trained models despite the bitwidth constrained to at least one bit, turning it into a Deep Set which explains its vastly different resource usage. As the HGQ trained models all have the same target EBOPs, the resource usage does not increase as happens with uniformly quantized models such as the Deep Sets trained in QKeras.

\begin{table}[h!]
    \centering
    \begin{adjustbox}{width=0.90\textwidth}

        \begin{tabular}{lcccccc}
            \toprule
            Model                                   & Particles & Acc. (\%) & Latn. (ns) & LUT (k) & II (clk) & DSP   \\
            \midrule
            Multi-Head Attention                    & 8         & 66.3      & 104         & 246    & 1        & 0   \\
            Multi-Head Attention                    & 16        & 72.3      & 98         & 279     & 1        & 0   \\
            Multi-Head Attention                    & 32        & 77.0      & 83         & 180     & 1        & 0   \\
            Multi-Head Attention                    & 64        & 77.9      & 44         & 47      & 1        & 0     \\
            \midrule
            Linformer                               & 8         & 66.3      & 110        & 230     & 1        & 0   \\
            Linformer                               & 16        & 72.8      & 103        & 246     & 1        & 0   \\
            Linformer                               & 32        & 78.4      & 140        & 267     & 1        & 0   \\
            Linformer                               & 64        & 79.8      & 78         & 202     & 1        & 0   \\
            \midrule
            Deep Sets (HGQ)                         & 8         & 64.7      & 49         & 177     & 1        & 0     \\
            Deep Sets (HGQ)                         & 16        & 70.1      & 53         & 205     & 1        & 0     \\
            Deep Sets (HGQ)                         & 32        & 77.4      & 53         & 256     & 1        & 0     \\
            Deep Sets (HGQ)                         & 64        & 79.4      & 44         & 191     & 1        & 0     \\
            \midrule
            MLP Mixer~\cite{mlpm-fpga}              & 16        & 71.7      & 68         & 75      & 1        & 0     \\
            MLP Mixer~\cite{mlpm-fpga}              & 32        & 78.0      & 62         & 63      & 1        & 0     \\
            MLP Mixer~\cite{mlpm-fpga}              & 64        & 79.7      & 72         & 159     & 1        & 0     \\
            \midrule
            Deep Sets (QKeras)~\cite{deepsetqkeras} & 8         & 64.0      & 95         & 386     & 3        & 626   \\
            Deep Sets (QKeras)~\cite{deepsetqkeras} & 16        & 69.4      & 115        & 747     & 3        & 555   \\
            Deep Sets (QKeras)~\cite{deepsetqkeras} & 32        & 75.9      & 130        & 903     & 2        & 434   \\
            \midrule
            Deep Sets M (QKeras)~\cite{nas-fpga}    & 8         & 65.1      & 110        & 130     & 3        & 548   \\
            Deep Sets L (QKeras)~\cite{nas-fpga}    & 8         & 66.6      & 135        & 337     & 3        & 2,458 \\
            \bottomrule
        \end{tabular}
    \end{adjustbox}

    \vspace{0.25cm}

    \caption{Performance comparison of the Transformer, Linformer, and other models. The table shows the FPGA implementation details in latency, look-up tables (LUT) used, initiation interval (II) in clock cycles and digital signal processing (DSP) blocks used. We show that the Linformers are able to achieve state-of-the-art accuracy while maintaining a reasonable low latency and realistic resource utilization for FPGA deployment.}
    \label{tab:comparison}
\end{table}

\section{Conclusion and Future Work}
\label{sec:conclusion}

This work introduces real-time transformer models with MHA and linear attention mechanisms with practical applications in particle physics. Of the two types of attention, Linformer is able to perform well over all input lengths evaluated, while MHA has lower performance due to the quadratic complexity leading to limited viability for FPGA deployment. All evaluated models achieve the required $\mathcal{O}(100)$ nanosecond latency with superior performance compared to the baseline models. Our code is available as part of the HGQ and \texttt{hls4ml} packages. In the future, we are looking to apply the transformer implementations in other real-time tasks in high-energy physics, such as jet tagging in high-pileup environment, particle reconstruction, event reconstruction and as a building block for a foundation model.

\FloatBarrier

\section{Acknowledgements}
\label{sec:acknowledgements}

Work done by Imperial College is funded by the Science and Technology Facilities Council (STFC) grant ST/W000636/1 and EPSRC (grant numbers UKRI256, EP/V028251/1, EP/N031768/1, EP/S030069/1, and EP/X036006/1).
B.M. acknowledges the support of Schmidt Sciences. A.G., and J.N. are supported by the DOE Office of Science, Award No. DE-SC0023524, FermiForward Discovery Group, LLC under Contract No. 89243024CSC000002 with the U.S. Department of Energy, Office of Science, Office of High Energy Physics, LDRD L2024-066-1, Fermilab, DOE Office of Science, Office of High Energy Physics ``Designing efficient edge AI with physics phenomena'' Project (DE-FOA-0002705), DOE Office of Science, Office of Advanced Scientific Computing Research under the ``Real-time Data Reduction Codesign at the Extreme Edge for Science'' Project (DE-FOA-0002501).

\FloatBarrier

\bibliographystyle{ieeetr}
\bibliography{bibliography}

\end{document}